\documentclass[12pt]{article}
\def\psl{\hbox{/\kern-.5800em$p$}}

\def\gappeq{\mathrel{\rlap {\raise.5ex\hbox{$>$}}
{\lower.5ex\hbox{$\sim$}}}}
\def\lappeq{\mathrel{\rlap{\raise.5ex\hbox{$<$}}
{\lower.5ex\hbox{$\sim$}}}}
\begin{document}
\pagestyle{empty}
\begin{flushright}
UAB-TH-521\\
March 2002
\end{flushright}
\vspace*{5mm}

\begin{center}
{\Large\bf A St{\"u}ckelberg formalism for the gravitino}

{\Large\bf  from warped extra dimensions}\\
\vspace{1.0cm}

{Tony Gherghetta$^a$ and Alex Pomarol$^{b}$}\\
\vspace{.5cm}
{\it\small {$^{a}$School of Physics \& Astronomy,
University of Minnesota,\\
Minneapolis, MN 55455, USA}}\\
{\it\small {$^{b}$ IFAE, Universitat Aut{\`o}noma de Barcelona,\\
E-08193 Barcelona, Spain}}\\
\vspace{.4cm}
\end{center}

\vspace{1cm}
\begin{abstract}

We consider supersymmetric theories with a warped extra dimension 
where supersymmetry is broken by boundary conditions that preserve 
an $R$-symmetry. It is shown that this supersymmetry breaking
mechanism naturally invokes the St{\"u}ckelberg formalism 
for the gravitino in order to give a four-dimensional 
theory with a smooth massless limit.

\end{abstract}

\vfill
\begin{flushleft}
\end{flushleft}
\eject
\pagestyle{empty}
\setcounter{page}{1}
\setcounter{footnote}{0}
\pagestyle{plain}

%============================================================

{\bf 1.}
The idea of the unification of forces makes us believe
that the Standard Model (SM) is replaced, at high energies,
by a more fundamental theory with larger symmetries.
Two interesting possibilities are grand unified theories and 
supersymmetric theories.
Of course, the extra symmetries of the fundamental theory 
must be broken, by some  mechanism, down to the SM symmetry group.
Extra dimensions  allow for  new possibilities to break  
the symmetries
of the more fundamental theory down to the Standard Model.
The idea is that these  symmetries could be realized in the higher 
dimensional theory but not in four dimensions. It is during 
the process of compactification that the symmetries are broken.
Therefore it is interesting to analyze the breaking of symmetries
by boundary conditions and to study its consistency.

In this Letter we want to study the fate of gravitinos in five-dimensional
(5d) warped spaces when supersymmetry is broken by boundary conditions.
It has been shown~\cite{gp2} that in these theories 
an $R$-symmetry (which is a chiral symmetry) can be preserved 
and, as a consequence, there are 
two light gravitinos, $\Psi_{\mu\, L}$ and $\Psi_{\mu\, R}$
($\Psi_{\mu\, L,R}=\frac{1}{2}(1\pm\gamma_5)\Psi_\mu$),
that combine to form a Dirac field. 
If the effective scale on the boundary where 
supersymmetry is broken is the TeV scale, then the gravitino mass
is of order $10^{-3}$ eV (instead of order the TeV scale 
as would be the case for a flat 5d space).
One of the gravitinos, $\Psi_{\mu\, L}$, is weakly coupled with
$1/M_P$-suppressed couplings, and is similar to the usual 
four-dimensional (4d) supersymmetric theories with 
supersymmetry broken at the TeV scale.
Therefore its presence is not unusual. However, the presence 
of another light gravitino (due to the  $R$-symmetry),
$\Psi_{\mu\, R}$,  is different from ordinary scenarios.
Furthermore, it has a coupling suppressed only by the supersymmetry 
breaking scale 1/TeV (which will be explained later why this is the case).
This raises the following apparent inconsistency.
Since this gravitino is coupled to a nonconserved
current $J_\mu$  (because supersymmetry is broken) its
contribution to an amplitude $\cal M$ is
\begin{equation}
     {\cal M}= {\bar J}^\mu P_{\mu\nu} J^\nu\, ,
\label{ampli}
\end{equation}
where $P_{\mu\nu}$ is the propagator of a  massive gravitino
\begin{equation}
    P_{\mu\nu}=\frac{1}{p^2+m^2}\Bigg[\Big(\eta_{\mu\nu}
       +\frac{p_\mu p_\nu}{m^2}\Big)(-i\psl-m)-\frac{1}{3}
       \Big(\gamma_\mu+i\frac{p_\mu}{m}\Big)(i\psl-m) 
       \Big(\gamma_\nu+i\frac{p_\nu}{m}\Big)\Bigg]\, .
\end{equation}
We are assuming a Minkowski metric $\eta_{\mu\nu}=(-1,1,1,1)$.
Notice that this amplitude diverges in the limit $m\rightarrow 0$.
In particular, for the coupling of the gravitino $\Psi_{\mu, R}$ to
the photon and photino via the fermionic current
\begin{equation}
     {\bar J}^\mu = \frac{g}{2\sqrt{2}M} \bar{\widetilde{\lambda}} 
     \gamma^\mu \gamma^{\nu\rho}  F_{\nu\rho}(1-\gamma_5)\, ,
\label{current}
\end{equation}
we can see, using dimensional analysis, that the amplitude ${\cal M}$
grows with the energy $E$ as
\begin{equation}
    {\cal M}\sim \frac{E^4}{m^2 M^2}\, .
\label{ampliE}
\end{equation}
Clearly, this amplitude will overwhelm unitarity bounds at energies 
$E\gappeq\sqrt{M m}$. 
For $M\sim M_P$ this consistently occurs at energies $E\gappeq $ TeV, and
there are no problems in the effective theory. 
However, for $M\sim$ TeV, this implies that the 
theory is not consistent at energies $E\gappeq 0.1$ MeV, which is much 
below the effective scale which is at the TeV. We will see how  
this apparent inconsistency is dealt with in the compactification 
of a higher-dimensional warped theory.

Before we consider supersymmetry and warped geometry it is
instructive to consider a simpler example where similar features
occur. In fact the above situation is similar to the case of a massive 
spin-1 particle coupled to a nonconserved current. 
The Lagrangian is given by
\begin{equation}
\label{abellag}
    {\cal L} = -\frac{1}{4} F_{\mu\nu} F^{\mu\nu}
             -\frac{1}{2} m^2 A_\mu A^\mu + A_\mu J^\mu\, ,
\end{equation}
where $J_\mu$ is the vector current containing the matter fermions.
The propagator of a massive spin-1 particle is given by
\begin{equation}
    \frac{1}{p^2+m^2}\left[\eta_{\mu\nu}+\frac{p_\mu p_\nu}{m^2}\right]\, ,
\end{equation}
and becomes singular in the  limit $m\rightarrow 0$.
Therefore the gauge boson contribution to the four-fermion amplitude 
grows with energy like $g^2E^2/m^2$, where $g$ is the gauge coupling, 
and will saturate the unitarity bounds at energies $E\sim m/g$.
In this case the procedure 
for achieving a smooth limit $m\rightarrow 0$ is known.
If we add a scalar field $\Phi$ with negative kinetic term
\begin{equation}
\label{stucksca}
  {\cal L}_\Phi =\frac{1}{2}\partial_\mu\Phi\partial^\mu\Phi
             +\frac{1}{m} \Phi\partial_\mu J^\mu\, ,
\end{equation}
then the four-fermion amplitude will receive an extra contribution 
mediated by $\Phi$ that cancels the singular term 
$\frac{p_\mu p_\nu}{m^2}$ coming from the massive gauge boson.
This can also be explicitly seen by performing the shift 
$A_\mu \rightarrow A_\mu +\partial_\mu\Phi/m$ in the Lagrangian  
(\ref{abellag}), and together with (\ref{stucksca}) leads to the
combined Lagrangian
\begin{equation}
\label{stuckscab}
    {\cal L} = -\frac{1}{4} F_{\mu\nu} F^{\mu\nu}
             -\frac{1}{2} m^2 A_\mu A^\mu + A_\mu J^\mu
             +m\Phi\partial_\mu A^\mu\, .
\end{equation}
In this form the field $\Phi$ appears as a Lagrange multiplier
which can be integrated out. This is achieved  by introducing a 
term $-\frac{\mu^2}{2}\Phi^2$ in the Lagrangian, performing a Gaussian
integration, and taking the limit $\mu\rightarrow 0$.
This finally leads to the Lagrangian
\begin{equation}
    {\cal L} = -\frac{1}{4} F_{\mu\nu} F^{\mu\nu}
             -\frac{1}{2} m^2 A_\mu A^\mu + A_\mu J^\mu
             +\frac{1}{2\alpha} (\partial\cdot A)^2\, ,
\end{equation}
with $\alpha=\mu^2/m^2\rightarrow 0$.
This is the Lagrangian for a massive vector field and
in the massless limit $(m=0)$ 
the last term corresponds to the Landau gauge-fixing term.  
The propagator now becomes
\begin{equation}    
   \frac{1}{p^2+m^2}\Big[\eta_{\mu\nu}-\frac{p_\mu p_\nu}{p^2}\Big]\, ,
\end{equation}
and there is no singular term in the limit $m\rightarrow 0$.
This is known as the St{\"u}ckelberg formalism~\cite{iz}~\footnote{
Notice that this is different from a non-linear realization of the
gauge symmetry, which in the U(1) case, is sometimes also called 
the St{\"u}ckelberg formalism.}.
The field $\Phi$ has the same couplings  
as a Nambu-Goldstone boson of the global U(1) 
but its kinetic term has an opposite sign. Therefore it
acts like a ``ghost'' (with negative energy and negative norm).
In the massless limit ($m\rightarrow 0$) and gauge-symmetric limit 
($\partial_\mu J^\mu=0$), $\Phi$ cancels the extra degree of freedom of 
the massive spin-1 boson leaving only the two degrees of freedom
of the massless gauge boson

This particular way of achieving a smooth $m\rightarrow 0$ limit 
using the St{\"u}ckelberg formalism is, as we shall discuss below,
the same way that warped higher dimensional theories recover a  
smooth  $m\rightarrow 0$ limit in the gravitino contribution to
amplitudes.

{\bf 2.} 
Let us consider a 5d supersymmetric theory compactified 
on the orbifold $S^1/Z_2$, where the two boundaries are located 
at $y^\ast=0$ and $y^\ast=\pi R$.
Fields living in the 5d bulk  must have definite boundary conditions.
For fermions, we can assume that
\begin{equation}
     \psi_{L,R}(0)=\pm\psi_{L,R}(0)\ ,
\ \ \ \ \
     \psi_{L,R}(\pi R)=\pm\psi_{L,R}(\pi R)\, .
\label{bc}
\end{equation}
If their supersymmetric bosonic partners have the
same boundary condition, the  theory
will remain supersymmetric after compactification.

Let us consider the effect of changing  the boundary conditions at
$y^\ast =\pi R$ from those at $y^\ast =0$ but only for the fermions.
This means that the 5d fermions of the theory satisfy
\begin{equation}
     \psi_{L,R}(0)=\pm\psi_{L,R}(0)\ ,
\ \ \ \ \
     \psi_{L,R}(\pi R)=\mp\psi_{L,R}(\pi R)\, .
\label{tbc}
\end{equation}
These boundary conditions break supersymmetry but 
preserve a U(1)$_R$ symmetry.
This mechanism is usually called 
Scherk-Schwarz supersymmetry breaking~\cite{ss}. 
In a flat 5d space the action 
for the gravitino $\Psi_M$, $(M=\mu,5)$, is given by
\begin{equation}
\label{action}
    S_5 =-\int d^5x  \sqrt{-g}\,  \Bigg[i
    \bar\Psi_M\gamma^{MNP}D_N\Psi_{P}+\left[i\bar J^M\Psi_{M}
+h.c.\right]\Bigg]\, ,
\end{equation}
and has an $N=2$  supersymmetry. If we impose
the  boundary conditions of Eq.~(\ref{tbc}) then
the $N=2$ supersymmetry is spontaneously broken to $N=0$.
In other words,
the $N=2$ supersymmetry is still realized but only nonlinearly.
After compactification all the gravitinos are massive. 
In particular, the lowest modes consist of two gravitinos,
$\Psi_{\mu\, L}$ and  $\Psi_{\mu\, R}$, with a 
mass $m=1/(2R)$. The components $\Psi_{5\, L}$ and $\Psi_{5\, R}$
play the role of Goldstinos~\cite{flat}, and can be eliminated 
by choosing the unitary gauge 
\begin{equation}
    \Psi_{\mu\, L}\rightarrow\Psi_{\mu\,L}
      -\frac{1}{m}\partial_\mu\Psi_{5\, L}\ , \ \ \ \ \ \ \ 
    \Psi_{\mu\, R}\rightarrow\Psi_{\mu\,R}
      +\frac{1}{m}\partial_\mu\Psi_{5\, R}\, . 
\label{redefi}
\end{equation}
The spectrum of this theory preserves an $R$-symmetry.

For warped spaces the situation is different.  Let us consider
the metric 
\begin{equation}
   ds^2 = a^2(y)\eta_{\mu\nu} dx^\mu dx^\nu + dy^2\, .
\label{metric}
\end{equation}
Although we believe that our results are valid for a general warp 
factor $a(y)$, for concreteness, 
we will focus on the Randall-Sundrum (RS) scenario~\cite{rs},
where $a(y)=e^{-k|y|}$, and $1/k$ is the curvature radius which is 
assumed to be of order the fundamental scale.
The 5d gravitino action is given by~\cite{susyrs}
\begin{equation}
\label{actionads}
    S_5 = -\int d^5x\sqrt{-g}\, \left[i
    \bar\Psi_M\gamma^{MNP}D_N\Psi_{P}
    +i\frac{3}{2}\frac{a^\prime}{a} \bar\Psi_M\gamma^{MN}
    \Psi_{N}+\left(i\bar J^M  \Psi_{M}+h.c.\right)\right]\, .
\end{equation}
Let us impose the boundary conditions Eq.~(\ref{tbc}) on the gravitino
wavefunction. This leads to the zero-mode gravitino wavefunctions
\begin{equation}
      f_L(y) \simeq \sqrt{2\pi k R}\,\,
       e^{-\frac{1}{2} k|y|} (e^{4 k(|y|-\pi R)}-1)\, ,
\end{equation}
and
\begin{equation}
      f_R(y) \simeq 2\sqrt{\pi k R}\,\,
       e^{\frac{k}{2}(|y|-4\pi R)} (e^{2 k|y|} -1)\, .
\end{equation}
Consequently, we see that the gravitino zero-mode
$\Psi_{\mu\,L}$ is localized on the 
boundary at $y^*=0$  (where the effective scale is $M_P$),
while $\Psi_{\mu\,R}$ is localized on the boundary at 
$y^*=\pi R$ (where the effective scale is TeV).
This will lead to a 4d theory with the features described in the
introductory remarks.

As in the 5d flat space case, we would like to know whether 
the 4d $N=1$ supersymmetry is nonlinearly realized by trying 
to eliminate the fields $\Psi_{5\, L,R}$. 
If we attempt to go to the unitary gauge, 
using the equivalent redefinitions for the 5d warped space (which
are given in Eq.~(12) of Ref.~\cite{gp2}), we find that $\Psi_{5L}$
is eliminated but that contrary to the flat case, the field 
$\Psi_{5R}$ does not disappear from the theory. 
Instead, we find that the zero-mode gravitino 4d Lagrangian is given by
\begin{eqnarray}
  {\cal L}_4 &=&-i\bar\Psi_{\mu\,L}
  \gamma^{\mu\nu\rho}\partial_\nu\Psi_{\rho\, L} -i\bar\Psi_{\mu\,R}
  \gamma^{\mu\nu\rho}\partial_\nu\Psi_{\rho\, R}\nonumber\\
  &+&\left[i m\bar\Psi_{\mu\, L}\gamma^{\mu\nu}\Psi_{\nu\,R}
  +i{\bar J}^\mu \Psi_{\mu\,R} +
  i\sqrt{\frac{3}{2}}m \bar\Psi_{\mu\, R}\gamma^\mu\Psi+h.c.\right]\, , 
\label{lagra}
\end{eqnarray}
where $m\simeq \sqrt{8}ke^{-2k\pi R}$ is the gravitino mass, and 
$\Psi$ is a linear combination of the Kaluza-Klein fields of $\Psi_5$ 
\begin{equation}
    \Psi(x) \simeq \sqrt{12\pi k R}\,\, e^{-\frac{1}{2}\pi k R}
              \sum_{n=0}^{\infty} \frac{f_R^{(n)}(\pi R)}{m_n} 
            \Psi_{5R}^{(n)}(x)\,.
\end{equation}
Notice that in Eq.~(\ref{lagra}) we have only considered the current 
to be coupled to $\Psi_{\mu\, R}$, and have neglected the couplings 
to $\Psi_{\mu\, L}$ since they are suppressed by $1/M_P$. 
As we discussed earlier, the presence of a light gravitino 
coupled to the nonconserved current of Eq.~(\ref{current})  
leads to a  photon-photino  scattering amplitude
\footnote{Notice that using Eq.~(\ref{current}) we have 
${\bar J}\cdot \gamma= \gamma\cdot J =0$.}
\begin{equation}
    {\cal M}_{3/2}={\bar J}^\mu\frac{(-i\psl)}{p^2+m^2}
      \left[\eta_{\mu\nu}+\frac{2}{3}\frac{p_\mu p_\nu}{m^2}\right]J^\nu\, ,
\label{ampligrav}
\end{equation}
that diverges in the limit $m\rightarrow 0$. 
However, the interesting feature in Eq.~(\ref{lagra}) is that
the field $\Psi$ appears as a Lagrange multiplier in an analogous 
way to the scalar field $\Phi$ in Eq.~(\ref{stuckscab}). This suggests 
that the fermion $\Psi$ will play a similar role to the scalar field 
$\Phi$ in recovering the smooth limit $m\rightarrow 0$. In fact, this 
is exactly what happens. If we perform the following redefinitions
\begin{eqnarray}
      \Psi_{\mu\, L}&\rightarrow&\Psi_{\mu\,L}
       -\frac{1}{\sqrt{6}}\gamma_\mu\Psi\, ,\\
       \Psi_{\mu\, R}&\rightarrow&\Psi_{\mu\,R}
       -\sqrt{\frac{2}{3}}\frac{1}{m}\partial_\mu\Psi\, ,
\end{eqnarray}
then the gravitino and St{\"u}ckelberg fermion, $\Psi$, decouple from
each other in the Lagrangian (\ref{lagra}), 
and leads to the separate Lagrangians
\begin{eqnarray}
    \nonumber \\
    {\cal L}_{3/2}&=&-i\bar\Psi_{\mu\,L}\gamma^{\mu\nu\rho}
    \partial_\nu\Psi_{\rho\, L} -i\bar\Psi_{\mu\,R}\gamma^{\mu\nu\rho}
    \partial_\nu\Psi_{\rho\, R}\nonumber\\
     &&\qquad\qquad\qquad\qquad+\big[i m\bar\Psi_{\mu\, L}
     \gamma^{\mu\nu}\Psi_{\nu\,R}+i {\bar J}^\mu\Psi_{\mu\, R}
     +h.c.\big]~,
    \label{gravi32}\\
    {\cal L}_{\Psi}&=&i\bar\Psi\gamma^\mu\partial_\mu\Psi
     +\left[i\sqrt\frac{2}{3}\frac{1}{m} \partial_\mu {\bar J}^\mu\Psi
    +h.c.\right]\, .
\label{stuck}
\end{eqnarray}
The Lagrangian (\ref{stuck}) is the analogue of Eq.~(\ref{stucksca}) in
the massive vector field example. We now see from Eq.~(\ref{stuck})
that the fermion $\Psi$ gives a new contribution to the amplitude of
Eq.~(\ref{ampligrav}), namely
\begin{equation}
    {\cal M}_\Psi= {\bar J}^\mu i\frac{2}{3}\frac{p_\mu p_\nu}{m^2}
     \frac{\psl}{p^2}J^\nu\, .
\label{sucon}
\end{equation}
Thus, combining the two contributions (\ref{ampligrav}) and (\ref{sucon})
we see that the divergent piece of Eq.~(\ref{ampligrav}) is precisely
cancelled leading to the combined total amplitude
\begin{equation}
  {\cal M}={\cal M}_{3/2}+{\cal M}_{\Psi}
       ={\bar J}^\mu\frac{(-i\psl)}{p^2+m^2}
      \left[\eta_{\mu\nu}-\frac{2}{3}\frac{p_\mu p_\nu}{p^2}\right]J^\nu\, ,
\label{totamp}
\end{equation}
which is smooth in the limit $m\rightarrow 0$. Again the similarity with
the vector field case is quite clear.

We have seen that the presence of the St{\"u}ckelberg fermion $\Psi$ 
guarantees that the amplitudes of the theory remain well behaved
at energies above the gravitino mass. Unlike the vector
field example where the St{\"u}ckelberg scalar was introduced by hand,
in the supersymmetric example the St{\"u}ckelberg fermion results
from the compactification of the warped geometry. 
However, in order that the above cancellation
takes place it is crucial that the
St{\"u}ckelberg fermion $\Psi$ has a kinetic term of opposite sign
compared to the other fermions in the theory --see Eq.~(\ref{stuck}).
Classically this is not a problem because $\Psi$ is a fermion, and 
its free energy is bounded from below (in fact this is a marked
improvement compared to the St{\"u}ckelberg scalar where classically
the energy is unbounded). Furthermore, at the quantum level, in  
processes where the gravitino and $\Psi$ appear off-shell, this 
will not present a problem either.
Nevertheless, the contribution of $\Psi$,
Eq.~(\ref{sucon}), has a pole at $p^2=0$ revealing that $\Psi$ can
be produced on-shell with negative probability. This is difficult
to interpret in quantum mechanics. The fermion $\Psi$  behaves like 
a ``ghost'' of negative norm.  
Although it is not clear how to make sense of this ghost in 
production processes, it is interesting to notice that
the Lagrangian of Eqs.~(\ref{gravi32}) and (\ref{stuck}) 
maintains its structure at the one-loop quantum level, {\it i.e.}, 
ultraviolet divergences can be absorbed by the counterterms of 
the tree-level couplings. This property is due to the fact 
that the theory consists of two sectors, namely, 
the Lagrangian of Eq.~(\ref{gravi32}) that corresponds 
to a spontaneously broken supersymmetric theory in the unitary gauge,
and the Lagrangian of Eq.~(\ref{stuck}) that contains the fermion 
$\Psi$ which behaves as a Goldstino of a global supersymmetry
but with  an opposite sign kinetic term. Thus, each sector is protected 
by its own supersymmetry (at least at the one-loop level). 
This property makes these theories 
theoretically interesting~\cite{GraviGhost}.

A similar situation to the above but for the case of gravitons instead 
of the gravitino is  present in the model of Ref.~\cite{grs}. 
In this theory the graviton is massive and a smooth massless limit 
is achieved due to the presence of a  ghost~\cite{ghost} 
that turns out to be the radion of the extra dimension~\cite{prz}.

{\bf 3.} We have seen that breaking supersymmetry by boundary 
conditions in warped geometries naturally invokes the St{\"u}ckelberg
formalism. Let us now discuss why supersymmetry breaking is qualitatively 
different in compact 5d theories with warped geometry compared 
to that encountered in 5d flat spaces.
The important thing to note is that in flat space supersymmetry  
is realized in the bulk of $S^1/Z_2$, and there is no need 
for boundary terms. Supersymmetry is always a good 
symmetry of the bulk independently of the boundary conditions of
the fields. If boundary conditions are different between bosons and 
fermions then supersymmetry is realized nonlinearly from a 4d
perspective. The 4d unitary gauge for the gravitinos corresponds 
to the 5d gauge $\Psi_5(x,y)=0$.

The situation changes in warped spaces. Supersymmetry in
warped spaces requires particular boundary couplings for the 
fields. For example, in the RS scenario with the boundary 
conditions (\ref{bc}), supersymmetry requires~\cite{susyrs}
a cosmological constant term on the two boundaries,
$\Lambda_{(0)}$ and $\Lambda_{(\pi R)}$ , related to the bulk cosmological
constant $\Lambda$ in the following way
\begin{equation}
\Lambda_{(0)}=-\Lambda_{(\pi R)}=-\Lambda/k\, .
\label{cc}
\end{equation}
These are precisely the conditions needed for the RS metric~\cite{rs}.
If we now change the boundary conditions of the bulk fermions
(but not the boundary couplings) then supersymmetry is no longer 
realized. There is an explicit breaking on the boundary.
Consequently, $\Psi_5(x,y)$ cannot be gauged away as in the flat case.
This explains the appearance of $\Psi_5$ in the spectrum. 
Let us be more explicit. If the fermions of the theory 
satisfy the boundary conditions (\ref{tbc}), then
the relation between the cosmological constants required by
supersymmetry is no longer given by Eq.~(\ref{cc}) but instead given by
\begin{equation}
\Lambda_{(0)}=\Lambda_{(\pi R)}=-\Lambda/k\, .
\label{cctbc}
\end{equation}
These relations, however, do not lead to the RS metric and in fact, 
do not even lead to a static configuration~\cite{bl}.
If we insist on obtaining the RS metric we must 
add an explicit supersymmetry breaking term of the form
\begin{equation}
    \int d^5x\, 2\frac{\Lambda}{k}\delta(y-\pi R)\,~.
\end{equation}
This term only breaks one of the two bulk supersymmetries.
Due to Eq.~(\ref{tbc}), the supersymmetry parametrized by
$\eta_{L}$ vanishes on the boundary at $y^*=\pi R$,
and is insensitive to terms on that boundary.
Only the supersymmetry parametrized by
$\eta_{R}$  is broken explicitly, and it is for this reason
that $\Psi_{5\, R}$  cannot be eliminated from the spectrum. 

The appearance of the St{\"u}ckelberg fermion seems to occur
in theories with (1) a localized gravitino at $y^*=0$, 
(2) supersymmetry broken at the opposite boundary at $y^*=\pi R$,
and (3)  $R$-symmetry invariance.
The first two conditions are necessary if one wants to explain
the hierarchy between $M_P$ and the TeV scale by a warp factor.
Therefore the only possibility to avoid the presence of 
the St{\"u}ckelberg fermion is to relax the third condition
and break the underlying $R$-symmetry. For example, this will be 
the case for any supersymmetry breaking mechanism that generates a 
Majorana mass for the gravitinos and gauginos on the $y^*=\pi R$
boundary. This will make all the  modes of the gravitino 
$\Psi_{\mu\, R}$ heavy (with masses of order TeV),
and the problem of unitarity discussed above will not arise 
(see Ref.~\cite{gp2,mp} for an example).
In this modified scenario the effective warped model at low energies 
contains only one light gravitino, and we recover an explicit realization 
of the usual scenarios considered in phenomenological
studies of superlight gravitinos~\cite{pheno}.

Let us finally comment on similar possibilities involving 
gauge symmetries. Either in flat or warped space the gauge symmetry 
can be realized in the bulk without the need for boundary terms. 
For example, in an Abelian gauge theory, the Lagrangian is given by 
\begin{equation}
\int d^4 x dy \sqrt{-g} \frac{1}{4}F_{MN}F^{MN}~, 
\end{equation}
which is invariant under the transformation 
$A_M\rightarrow A_M+\partial_M\theta$, irrespective of whether the space
is flat or warped. 
Therefore the breaking of the gauge symmetry by boundary conditions
is always nonlinearly realized with $A_5$ playing the role 
of the Nambu-Goldstone boson.

{\bf Acknowledgments:}
T.G. thanks the CERN Theory Division for hospitality where part
of this work was done.
A.P. was partially supported by the CICYT Research Project
AEN99-0766 and DURSI Research Project 2001-SGR-00188.

\end{document}